\documentclass[twocolumn]{aastex631}
\usepackage{booktabs}
\usepackage{array}
\usepackage{bm}
 \usepackage[slantedGreek]{newtxmath}
\usepackage{color}
\usepackage{subfigure}
\usepackage{graphicx}
\usepackage{figsize}
\usepackage{mathtools} 
\usepackage{url}
\usepackage{bookmark}

\begin{document}

  \title{PKS~J0805$-$0111: A Second Owens Valley Radio Observatory Blazar Showing Highly Significant Sinusoidal Radio Variability - The Tip of the Iceberg}


\author{P. V. de la Parra}
\affiliation{CePIA, Astronomy Department, Universidad de Concepci\'on,  Casilla 160-C, Concepci\'on, Chile}
\author{S. Kiehlmann}
\affiliation{Department of Physics and Institute of Theoretical and Computational Physics, University of Crete, 71003 Heraklion, Greece}
\author{P. Mr{\'o}z}
\affiliation{Astronomical Observatory, University of Warsaw, Al. Ujazdowskie 4, 00-478 Warszawa, Poland}
\author{A.C.S. Readhead}
\affiliation{Owens Valley Radio Observatory, California Institute of Technology,  Pasadena, CA 91125, USA}
\affiliation{Institute of Astrophysics, Foundation for Research and Technology-Hellas, GR-70013 Heraklion, Greece}
\author{A. Synani} 
\affiliation{Department of Physics and Institute of Theoretical and Computational Physics, University of Crete, 71003 Heraklion, Greece}
\affiliation{Institute of Astrophysics, Foundation for Research and Technology-Hellas, GR-70013 Heraklion, Greece}
\author{M. C.  Begelman}
\affiliation{JILA, University of Colorado and National Institute of Standards and Technology, 440 UCB, Boulder, CO 80309-0440, USA} 
\affiliation{Department of Astrophysical and Planetary Sciences, University of Colorado, 391 UCB, Boulder, CO 80309-0391, USA}
\author{R. D. Blandford}
\affiliation{Kavli Institute for Particle Astrophysics and Cosmology, Department of Physics,20
Stanford University, Stanford, CA 94305, USA}
\author{Y. Ding}
\affiliation{Cahill Center for Astronomy and Astrophysics, California Institute of Technology, Pasadena, CA 91125, USA}
\author{F. Harrison}
\affiliation{Cahill Center for Astronomy and Astrophysics, California Institute of Technology, Pasadena, CA 91125, USA}
\author{I. Liodakis}
\affiliation{NASA Marshall Space Flight Center, Huntsville, AL 35812, USA}
\affiliation{Institute of Astrophysics, Foundation for Research and Technology-Hellas, GR-70013 Heraklion, Greece}
\affiliation{Finnish Center for Astronomy with ESO, University of Turku, Vesilinnantie 5, FI-20014, Finland}
\author{W. Max-Moerbeck} 
\affiliation{Departamento de Astronomía, Universidad de Chile, Camino El Observatorio 1515, Las Condes, Santiago, Chile}
\author{V. Pavlidou} 
\affiliation{Department of Physics and Institute of Theoretical and Computational Physics, University of Crete, 71003 Heraklion, Greece}
\affiliation{Institute of Astrophysics, Foundation for Research and Technology-Hellas, GR-70013 Heraklion, Greece}
\author{R. Reeves}
\affiliation{CePIA, Astronomy Department, Universidad de Concepci\'on,  Casilla 160-C, Concepci\'on, Chile} 
\author{M. Vallisneri}
\affiliation{Jet Propulsion Laboratory, California Institute of Technology, 4800 Oak Grove Drive, Pasadena, CA 91109, USA}
\author{M.F. Aller}
\affiliation{Department of Astronomy, University of Michigan, 323 West Hall, 1085 S. University Avenue, Ann Arbor, MI 48109, USA}
\author{M. J. Graham}
\affiliation{Division of Physics, Mathematics, and Astronomy, California Institute of Technology, Pasadena, CA 91125, USA}
\author{T. Hovatta}
\affiliation{Finnish Centre for Astronomy with ESO (FINCA), University of Turku, FI-20014 University of Turku, Finland}
\affiliation{Aalto University Mets\"ahovi Radio Observatory,  Mets\"ahovintie 114, 02540 Kylm\"al\"a, Finland}
\author{C. R. Lawrence}
\affiliation{Jet Propulsion Laboratory, California Institute of Technology, 4800 Oak Grove Drive, Pasadena, CA 91109, USA}
\author{T.~J.~W.~Lazio}
\affiliation{Jet Propulsion Laboratory, California Institute of Technology, 4800 Oak Grove Drive, Pasadena, CA 91109, USA}
\author{A.A.~Mahabal}
\affiliation{Division of Physics, Mathematics, and Astronomy, California Institute of Technology, Pasadena, CA 91125, USA}
\affiliation{Center for Data Driven Discovery, California Institute of Technology, Pasadena, CA 91125, USA}
\author{B. Molina}
\affiliation{CePIA, Astronomy Department, Universidad de Concepci\'on,  Casilla 160-C, Concepci\'on, Chile}
\author{S. O'Neill}
\affiliation{Department of Physics, Princeton University, Jadwin Hall, Princeton,
08540, NJ, USA.}
\author{T. J. Pearson}
\affiliation{Owens Valley Radio Observatory, California Institute of Technology,  Pasadena, CA 91125, USA}
\author{V. Ravi}
\affiliation{Owens Valley Radio Observatory, California Institute of Technology,  Pasadena, CA 91125, USA} 
\author{K. Tassis} 
\affiliation{Department of Physics and Institute of Theoretical and Computational Physics, University of Crete, 71003 Heraklion, Greece}
\affiliation{Institute of Astrophysics, Foundation for Research and Technology-Hellas, GR-70013 Heraklion, Greece}
\author{J. A. Zensus}
\affiliation{Max-Planck-Institut f\"ur Radioastronomie, Auf dem H\"ugel 69, D-53121 Bonn, Germany}

\begin{abstract}
Owens Valley Radio Observatory (OVRO) observations of supermassive black hole binary (SMBHB) candidate PKS~2131$-$021 revealed, for the first time, six likely characteristics of the phenomenology exhibited by SMBHB in blazars, of which the most unexpected and critical is sinusoidal flux density variations. We have now identified a second blazar, PKS~J0805$-$0111, showing  significant sinusoidal variations, with an observed period that translates to $1.422 \pm 0.005$ yr in the rest frame of the  $z = 1.388$ object.
We generate $10^6$ simulated light curves to reproduce the radio variability characteristics of PKS~J0805$-$0111, and show that the global probability, considering the \textit{look-elsewhere effect}, indicates that the observed periodicity can be attributed to the red noise tail of the power spectral density, with a $p_0$ value of $7.8 \times 10^{-5}$ (i.e. 3.78$\sigma$). PKS J0805$-$0111  displays all six characteristics observed in PKS 2131$-$021. Taking into account the well-defined OVRO sample size, the false positive probability $\sim 0.22$, but the rare behavior makes this a strong SMBHB candidate.  The discovery of a second SMBHB candidate exhibiting these rare characteristics reveals that PKS~2131$-$021 is not a unique,  isolated case. With these two strong cases we are clearly seeing only the tip of the iceberg. We estimate that the number of SMBHB candidates amongst blazars $\sim$ 1 in 100.
\end{abstract}


%

\section{INTRODUCTION}

We are engaged in a  search  for statistically significant periodic and quasi-periodic oscillations (QPOs) in flux density amongst  the $\sim 1830$ blazars in the Owens Valley Radio Observatory (OVRO) 40\,m Telescope 15\,GHz monitoring program, of which 1158 form a complete well-defined sample \citep{Richards2011}.  
Compelling evidence of a stable periodicity in the blazar PKS~2131$-$021 from this program was recently presented by \citet[hereafter O22]{2022ApJ...926L..35O} who showed that it is highly unlikely to be due to random fluctuations due to the red noise in the variability spectrum. More recently \citet[hereafter K24]{PKS2131II} discovered coherent sinusoidal variations in PKS 2131$-$021 from 2.7 GHz to optical wavelengths. The very broadband coherent sinusoidal variations and the stability of the period over 47.9 years amount to compelling evidence, in our view, that PKS~2131$-$021 is a supermassive black hole binary (SMBHB). 

 \begin{figure*}[t!h]
   \centering
   \includegraphics[width=0.8\linewidth]{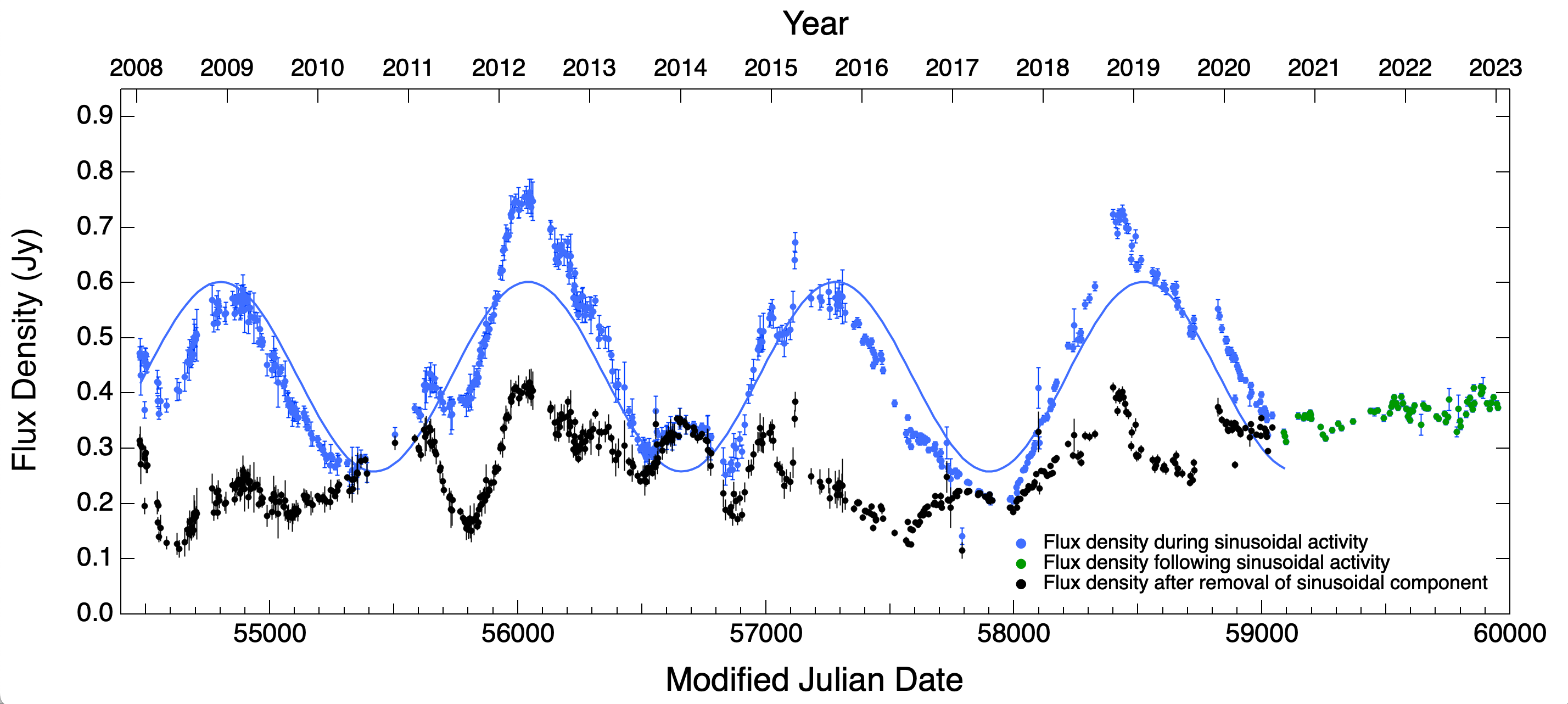}
      \caption{The light curve of PKS~J0805$-$0111. The 15\,GHz OVRO data are shown by the blue and green dots. The least-squares sine-wave fit to the data up to MJD~59041 is shown by the blue curve, the parameters of which are listed in Table \ref{tab:least_squares_fits}. The black dots show the residual randomly varying component of the OVRO light curve after subtracting the sinusoidal component. The periodicity in the light curve up to MJD~59041 has a global significance  of $3.5\sigma$, and is not likely to be simply a product of the red-noise tail in the PSD. }
         \label{plt:0805}
\end{figure*}

Based on their observations of PKS~2131$-$021, O22 drew attention to six likely characteristics of SMBHBs as revealed through their blazar light curves:
\vskip 6pt
\noindent
1. Strong sinusoidal flux-density variations. 
\vskip 6pt
\noindent
2. Long term stability of the period.
\vskip 6pt
\noindent
3. Rapid disappearance of the sinusoidal fluctuations in less than one period.
\vskip 6 pt
\noindent
4. Re-appearance of sinusoidal fluctuations in less than one period and with different amplitude.
\vskip 6pt
\noindent
5. The sinusoidal variations are in addition to the regular variations in the blazar; in other words, the mean flux density is elevated during the periods of the sinusoidal variations.
\vskip 6pt
\noindent
6. Variations in the period of $\sim 10\%$, or variations in phase within $\sim10$\% of a cycle.
\vskip 6pt
\noindent
PKS~J0805$-$0111 exhibits all six characteristics. \cite{2021RAA....21...75R} first drew attention to PKS~J0805$-$0111 as an interesting SMBHB candidate based on our 40\,m Telescope monitoring observations. We show that the  global $p$-value of  the periodic behavior in PKS~J0805$-$0111 being due to red noise tail random variations is small  ($p$-value = $7.8 \times 10^{-5}$, $3.78\sigma$), and therefore passes the $3\sigma$ threshold we have adopted for a strong SMBHB candidate (O22).

The source PKS J0805$-$0111 has been identified as a flat-spectrum radio quasar \citep{fsrq}, i.e., $\alpha > -0.5$, where $S \propto \nu ^\alpha$. The spectroscopic redshift of $z= 1.388$ was measured by \citet{Healey2008}.

As discussed in O22, blazars showing evidence of periodicities  should not be considered bona fide periodic or quasi-periodic oscillator (QPO) candidates unless it can be shown that there is a very low global probability ($p$-value $<1.3 \times 10^{-3}$, i.e. $ 3\sigma$) that the periodicity is due to red-noise tail fluctuations. This can be done through a careful analysis of the power spectral density (PSD) with simulations that reproduce both the observed PSD and the observed probability density function (PDF) seen in the light curve.

There are now several SMBHB candidates identified through a variety of methods (for a recent review see \citealp{Dorazio2023}).  Periodic variability has been systematically explored in large-scale optical surveys for active galactic nuclei (AGN, e.g., \citealp{Graham2015,Charisi2016}).  Apart from PKS 2131$-$021, the only other strong blazar candidates are OJ~287 (optical, \citealp{Sillanpa1988,2016ApJ...819L..37V,2021MNRAS.503.4400D}) and PG~1553+113 ($\gamma$-rays, \citealp{Ackermann2015}), and it should be noted that the latest predicted flare in OJ~287 was not observed \citep{Komossa2023,Valtonen2023}. In the radio band, while claims have been made for quasi-periodicities in several sources, PKS~2131$-$021 has been, until now, the only blazar showing clear sinusoidal variability, a signal which passes the global (i.e., the {\it look-elsewhere effect} corrected) $3\sigma$ significance threshold (O22). 

Therefore, PKS~J0805$-$0111 is only the second blazar to be identified as a high-probability SMBHB showing  sinusoidal variations in its radio light curve. This demonstrates that PKS~2131$-$021 is not a unique, isolated,  case.  Clearly, while this is an easily identifiable  class of AGN, it requires long-term radio monitoring observations in order to capture the 3~yr--5~yr periods of the candidates. In  this paper we argue that at least 1 in 1000 blazars shows strong sinusoidal variations over a large fraction of a 15-year light curve, and should therefore be considered  strong SMBHB candidates. Furthermore, we argue that, taking into account those blazars in which sinusoidal fluctuations are weaker, or absent,  for a significant fraction of the time, at least  1 in 100 blazars is an SMBHB candidate.

The radio observations of PKS~J0805$-$0111 are described in \S\ref{sec:radio}.   The analysis identifying PKS~J0805$-$0111 as a strong SMBHB candidate is presented in  \S\ref{sec:analysis}. The significance of periodicities discovered in this case and in OVRO radio light curve searches in  general is discussed in \S\ref{sec:significance}.  The effects of the sample size are discussed in \S \ref{sec:samplesize}. The fact that we are only identifying a small fraction of the blazars that are potential SMBHB candidates is considered in \S\ref{sec:iceberg}. Infrared and optical observations of PKS~J0805$-$0111 are described in \S\ref{sec:infrared}. The Hard X-ray and $\gamma-$ray emission from PKS J0805$-$0111 is discussed in \S\ref{sec:xand gamma}. The search for evidence of gravitational waves from PKS~J0805$-$0111 is discussed in \S\ref{sec:grav}.  Our conclusions are presented in \S\ref{sec:conclusions}.
  
  For consistency with our other papers, we assume the following cosmological parameters: $H_0 = 71$\, km\,s$^{-1}$\, Mpc$^{-1}$, $\Omega_{\rm m} = 0.27$, $\Omega_\Lambda = 0.73$  \citep{2009ApJS..180..330K}.  None of the conclusions would be changed were we to adopt the best model of the Planck Collaboration  \citep{2020AandA...641A...6P}. 

\section{Radio Observations - OVRO 40 m }\label{sec:radio}


The OVRO 40\,m telescope has been monitoring $\sim1830$ blazars at 15\,GHz since 2008 \citep{Richards2011}.  The telescope is equipped with a cryogenic pseudo-correlation receiver. The observing cadence is $\sim$1--2 measurements per week. Occasional gaps exist in the data due to weather conditions or hardware problems. The OVRO 40\,m Telescope receiver uses a dual-beam switching procedure to remove atmospheric contributions and background emission, as described by \textcolor{blue}{\citet{readhead1989}}.  At this radio frequency the observations are confusion-limited, due to the double-switching technique which combines three separate fields,  and the resulting flux density uncertainty   $\sim$3--4 mJy. The data reduction and calibration process used to produce the light curves is described in \textcolor{blue}{\cite{Richards2011}}.

 \section{Analysis of the Radio Light Curve of PKS J0805\texorpdfstring{$-$}{-}0111}\label{sec:analysis}

 The OVRO 40\,m Telescope 15\,GHz light curve of PKS~J0805$-$0111 is shown in Fig.~\ref{plt:0805}.  As in the case of PKS~2131$-$021 (O22), a clear sinusoidal variation dominates the variability for a significant fraction of the period of observation. The blue curve shows the least-squares sine-wave fit to the data up to MJD~59041,  with an observed period of $1240.7\pm 4.6$ days, after which the sinusoidal variations disappear.

Due to weather and hardware problems the observations are not equispaced in time. We require statistically robust estimates of the probabilities of various features in the observed light curve, so we need a light curve simulation procedure that  preserves all of the statistical and variation characteristics of the observed light curves, including the identical irregular sampling.  We can then simulate a large number of light curves for each blazar and use a Monte Carlo approach to estimate the probability of chance occurrence of various features in the light curve. This approach  tests, with the same sampling and observational noise as the real data, the null hypothesis that the observed features in the light curve are a result of red-noise processes.
This problem  has been considered in a number of papers \citep{1999PASP..111.1347W,2002MNRAS.332..231U,2010MNRAS.404..931E,2014MNRAS.445..437M}, which simulate  light curves with power spectral density (PSD)  of the same variability power law slope as that observed in the AGN. Such simulations can be produced with the algorithm introduced by \citet{1995A&A...300..707T}.
However, the underlying probability density function (PDF) of such simulations is Gaussian, and does not produce realistic light curves. Radio AGN light curves exhibit ``burst''-like events that can often yield long-tailed PDFs. This problem was addressed by \cite{2013MNRAS.433..907E} using a simple method to produce light curves that match both the PSD and the PDF of the observed light curves. 
As desired, the final artificial light curves have the best matching statistical and variability properties of the observed light curve under the assumption of a red-noise process with power-law spectrum, and are statistically (and visually) indistinguishable from the true light curves.
We are therefore confident that the significances we calculate are robust, having correctly taken into account the red-noise tail in the observed PSD.
The details of the simulations are presented in Appendix \ref{app:sim}.

\subsection{Generalized Lomb-Scargle Analysis of the Light Curve}

\begin{figure}[t]
   \centering
   \includegraphics[width=1.\linewidth]{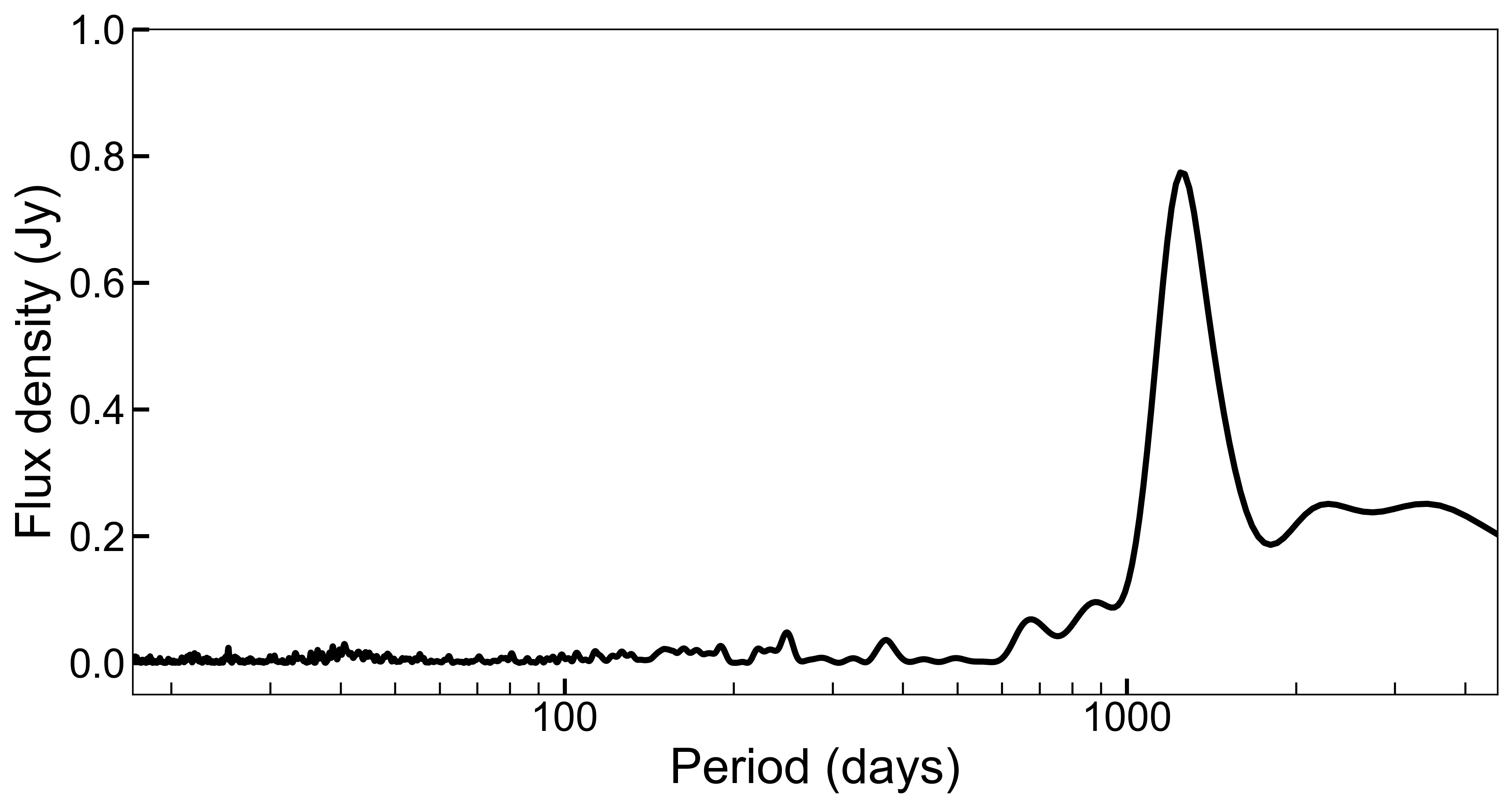}
   \caption{The generalized Lomb-Scargle (GLS) periodogram of PKS~J0805$-$0111 with power normalized to a range from 0 to 1, where 1 corresponds to a perfect fit, within the uncertainties, to a sine wave.}
   \label{plt:gls}
\end{figure}

\begin{figure}[t]
   \centering
   \includegraphics[width=1.\linewidth]{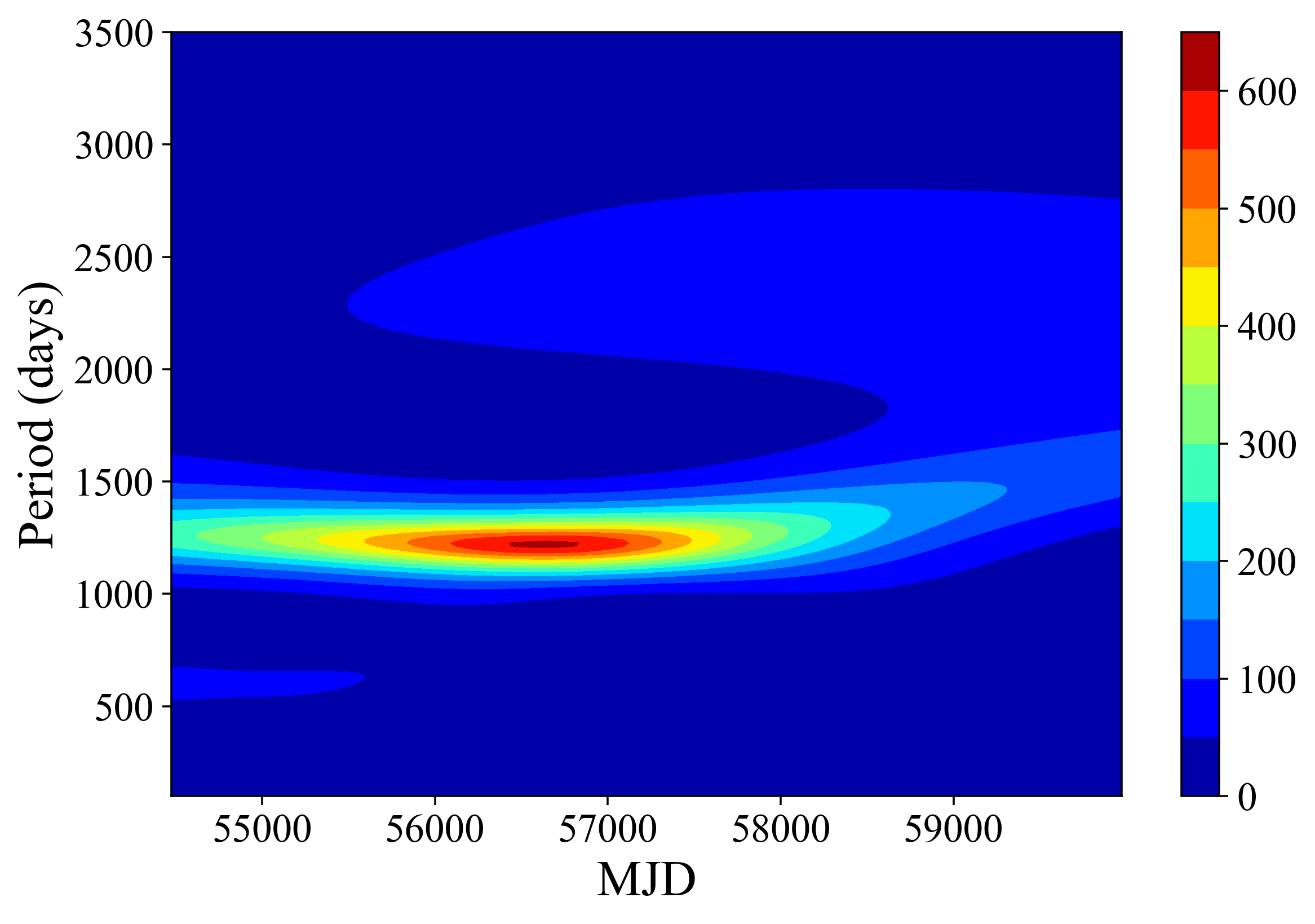}
   \caption{The WWZ analysis of PKS~J0805$-$0111}
   \label{plt:wwz}
\end{figure}

We carried out a Generalized Lomb-Scargle (GLS) periodogram analysis \citep{GLSP} of the light curve of PKS~J0805$-$0111, from MJD~54476 to  MJD~59041, shown by the blue points in Figure \ref{plt:0805}. This technique is useful for identifying periodicities in unevenly sampled data by minimizing least squares to fit sinusoids to the data.  The GLS periodogram is shown in Fig.~\ref{plt:gls}.  There is only one prominent peak at period $P=1245$ days, and power $\mathcal{P}=0.77$. We use $P$ and $\mathcal{P}$ to distinguish between period and power, respectively.

The frequency span of the periodogram is from $f_\mathrm{low} = 1/T$, where $T$ is the total observing time, to $f_\mathrm{high} = N/(2T)$, which corresponds to half of the average time interval between measurements, where $N$ is the total number of measurements. The power was calculated at evenly spaced frequency intervals in steps of $\Delta f = 1/(\zeta T)$. 
To evaluate the effect of different sampling intervals, we experimented with values of $\zeta$ = 5, 10, and 15. No differences were found in the significance analysis, so the value of $\zeta$ does not affect our conclusions. To maintain the highest resolution in our analysis we used $\zeta = 15$.

\begin{table*}
\centering
\caption{Results of the sine-wave least-squares fits}
\label{tab:least_squares_fits}
\begin{tabular}{lrrr}
\hline
Parameter               & $54476 < \mathrm{MJD} < $ 59041 & $54476<\mathrm{MJD}<56758$ & $56763<\mathrm{MJD}<59041 $\\
 & (full data) & (first two cycles) & (last two cycles) \\
\hline
\hline
$P$ (days)              & $1240.7 \pm 4.6$    & $1304.8 \pm 15.7$   & $1326.4 \pm 10.5$\\
$\phi_0$                & $-0.141 \pm 0.025$  & $0.269 \pm 0.082$  & $-0.525 \pm 0.043$\\
$A$ (Jy)                & $0.1709 \pm 0.0045$ & $0.1657 \pm 0.0059$ & $0.2000 \pm 0.0058$\\
$S_0$ (Jy)              & $0.4295 \pm 0.0031$ & $0.4255 \pm 0.0041$ & $0.4176 \pm 0.0040$\\
$\sigma_0$ (Jy)         & $0.0663 \pm 0.0022$ & $0.0646 \pm 0.0029$ & $0.0544 \pm 0.0028$\\
\hline
\end{tabular}
\end{table*}

\subsection{WWZ Analysis of the Lightcurve.}

The weighted wavelet Z-transform (WWZ) is a Z-transform \citep{ragazzini1952analysis} that uses wavelet analysis to detect time-dependent periodicities in data. We performed a WWZ analysis of the PKS~J0805$-$0111 lightcurve, with the results  shown in Fig.\ \ref{plt:wwz}. Following the approach presented in \cite{foster1996}, we used a 100-day time bin and tested over periods from 100 to 3500 days. Analysis beyond this range of parameters is not relevant to our conclusions. This analysis clearly shows the presence of a periodic oscillation that dominates, with a period $\sim$ 1250 days from the beginning of the lightcurve. This periodic signal remains stable in its period until around 59000 MJD, supporting the results obtained from the GLS periodogram.

\subsection{Sine Wave Analysis of the Light Curves.}\label{sec:sine}

We carried out a sine-wave least-squares fitting of the 15\,GHz light curve of PKS~J0805$-$0111 following the procedure described in O22, with the result that the period of the sinusoidal variation in PKS~J0805$-$0111 is $P=1240.7 \pm 4.6$ days, corresponding to a period in the rest frame of the blazar of $P=519.6 \pm 1.9$ days, at the redshift of $1.388$ \citep{Healey2008}. Results of the fit to the data from MJD~54476 to MJD~59041, when the sinusoidal variations disappeared,  are presented in the second column of Table~\ref{tab:least_squares_fits}.

\begin{figure}[!t]
   \centering
   \includegraphics[width=1\linewidth]{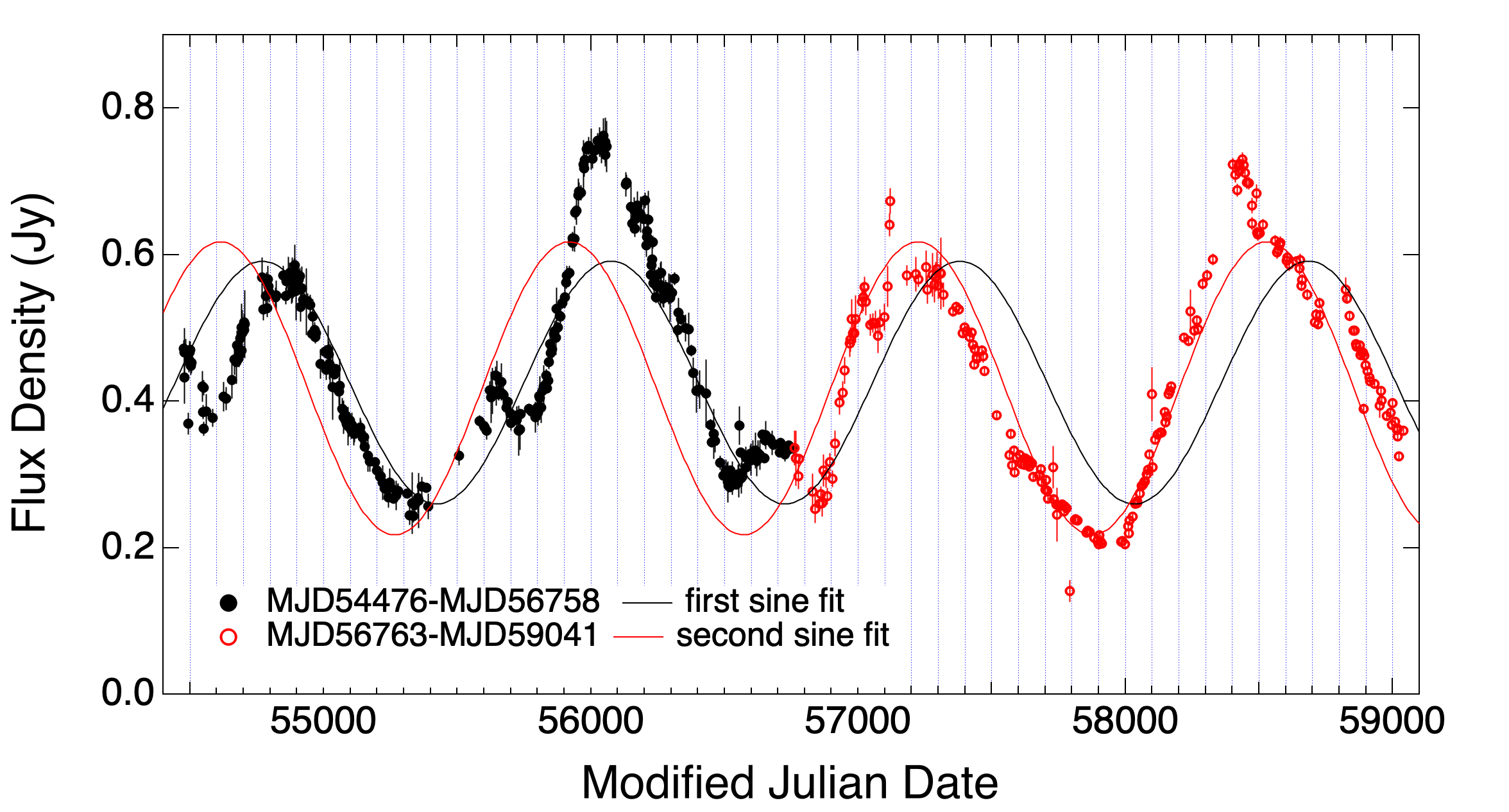}
      \caption{Sine-wave fits to the first and second halves of the light curve of PKS~J0805$-$0111 during the period showing sinusoidal fluctuations. The periods in the two halves are $1304.8\pm 15.7$ (black points and black curve); and $1326.4\pm 10.5$ (red points and red curve) }
         \label{plt:twosines}
\end{figure}

\subsubsection{Analysis of Two Halves of the Light Curve}\label{sec:sine0805}

\begin{figure*}[t]
   \centering
   \includegraphics[width=1.\linewidth]{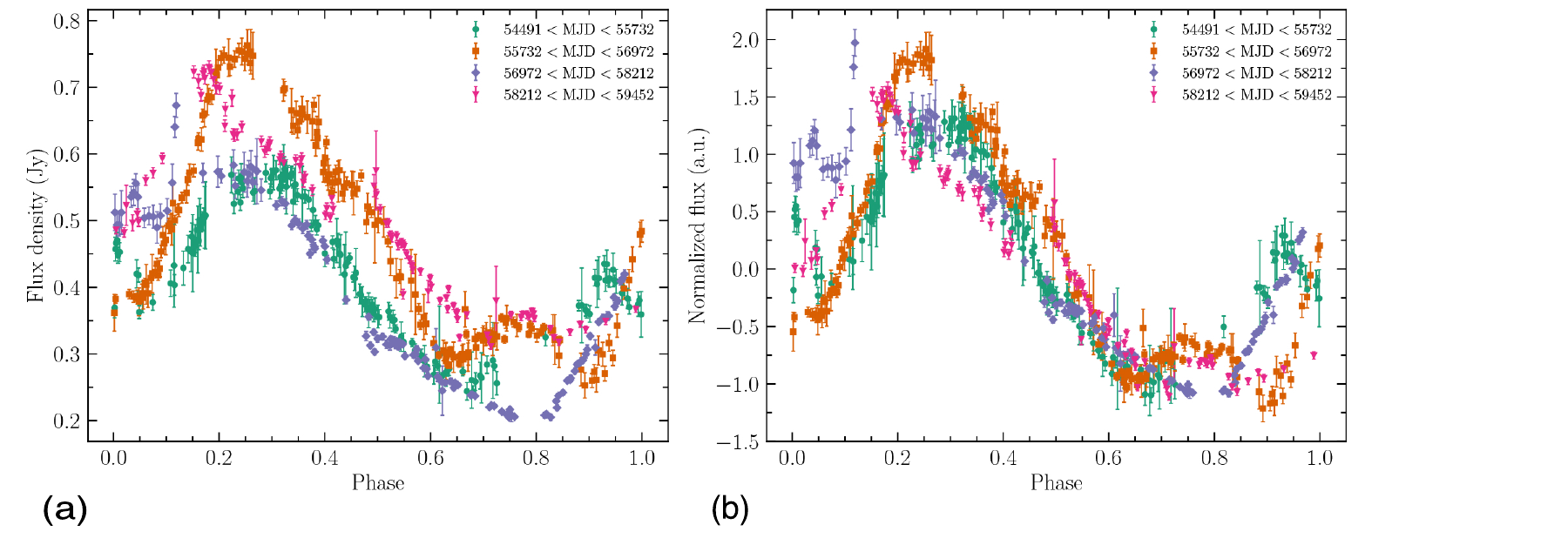}
      \caption{Light curve of PKS~J0805$-$0111 folded with a period of 1240.7 days. Note that in the final cycle the sinusoidal variation ceases at MJD 59041, so that the final portion of the light curve diverges from the other three curves. (a) The  light curves showing the observed flux densities in janskys. (b)  The data in each period normalized to the same offset (0) and amplitude (1): i.e. normalized flux density = (flux density - $S_0$) / A,
where A and $S_0$ were found using a least squares fit to the data. }
         \label{plt:0805phased}
\end{figure*}

\begin{figure*}[!ht]
   \centering
   \includegraphics[width=1.\linewidth]{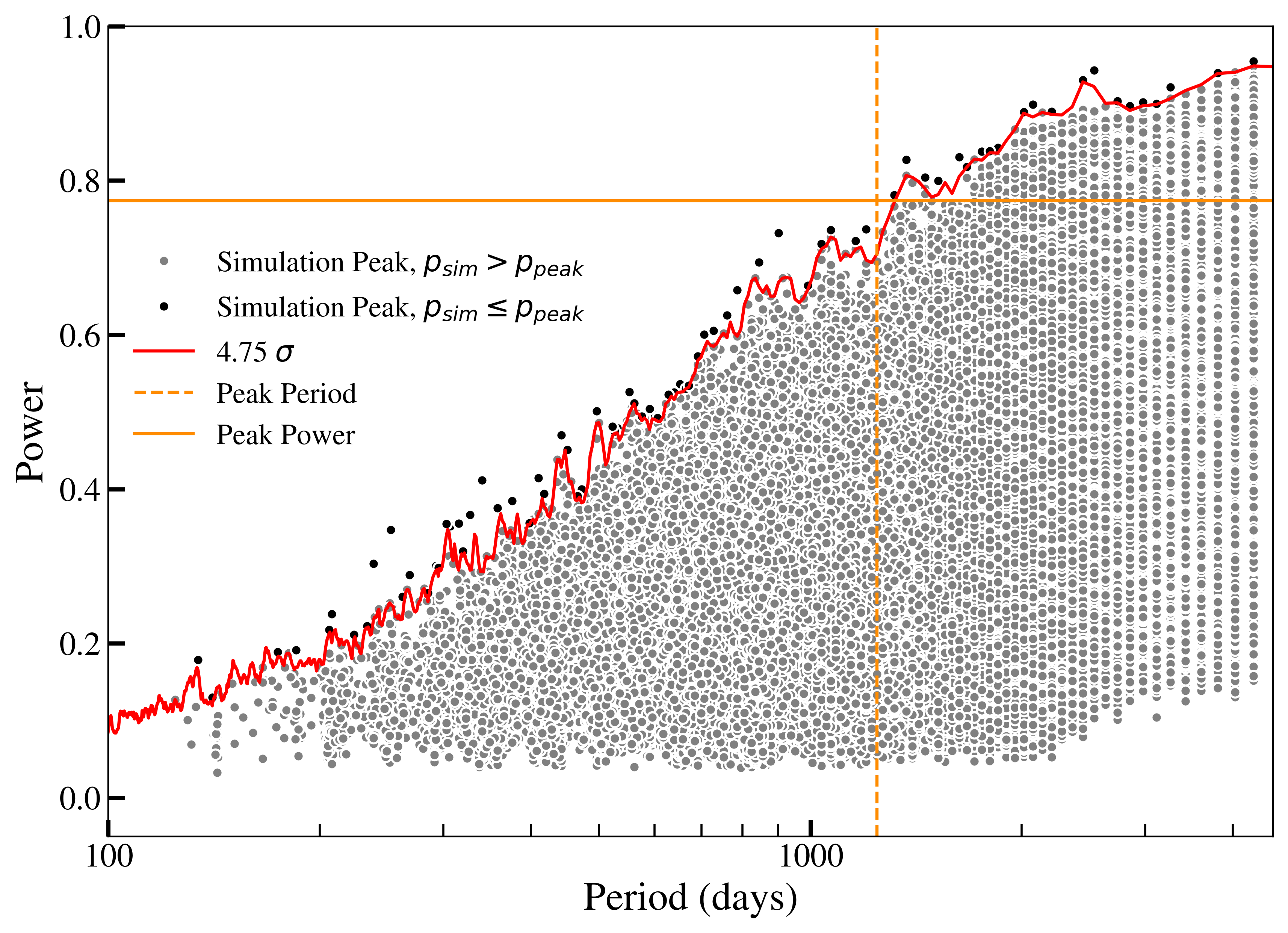}
      \caption{The generalized Lomb-Scargle peak power and period for each of the $1,000,000$  simulations is shown by the circles. Simulations where the local $p$-value $p_\mathrm{sim}$ is equal to or smaller than the local $p$-value of the observed GLSP peak $p_\mathrm{peak}$  are shown in black dots, whereas gray dots indicate $p_\mathrm{sim} > p_\mathrm{peak}$. The value of the period and power of the most significant peak in the GLS periodogram for PKS~J0805$-$0111 are shown by the vertical dashed and horizontal orange lines, respectively.  Note that the $4.75 \sigma $ line refers to the situation where the periods are known {\it a priori}. This sigma level is given by the local p-value calculated at the observed peak, $p_{peak} <1/1,000,000$. In the absence of such knowledge one has to use the global significance, which arises from calculating the fraction of black dots among all 1,000,000 dots. The global significance is considerably lower than what one would determine from the $4.5\sigma$- curve shown here (see O22). The quantization of the periods is due to the selection of the frequency resolution for the analysis, $\Delta f = 1/(\zeta T)$, with $\zeta= 15$.}
         \label{plt:glsp}
\end{figure*}

\begin{deluxetable*}{c@{\hskip 8mm}ccccc}
\tablecaption{GLS Probability and Significance Level from  Matched Red Noise Tail Simulations - Isolated Case}
\tablehead{Test& GLS $\mathcal{P}_{\rm peak}$ & Total& Number of simulations& $p$-value$^*$ & Significance$^*$\\
&(max=1)&Simulations&that pass test&&($\sigma$)}
\startdata
Global Significance& 0.77&1,000,000&78&$7.8 \times 10^{-5}$&$3.78$\\
\enddata
\tablecomments{  $\mathcal{P}$ is the GLS power. We count simulations at all periods with local $p$-values less than the local $p$-value of the peak in Fig.~\ref{plt:gls}. The period of the peak is $P_\textrm{peak}=1245 \pm 139$ days. $^*$ The significance values reported in this Table do \emph{not} take in to account the observed sample size of 1158 blazars  (see Sec.~\ref{sec:samplesize}).}
\label{tab:lsvalues}
\end{deluxetable*}

We have also carried out a sine-wave fitting analysis  of the first vs.\ the second half of the light curve of PKS~J0805$-$0111 during the period when the sinusoidal variation is present.  The fits are shown in Fig.~\ref{plt:twosines}. The periods of the two halves agree within the errors (1.3\%), but, as can be seen in Fig.~\ref{plt:twosines}, the phases differ significantly, by $\sim$12\% of the period.  This makes an interesting contrast to the situation where we fitted the two different halves of the OVRO 15\,GHz light curve of PKS~2131$-$021 (O22), in which we found that the \textit{periods} differed by $\sim$10\%.  
Perusal of the sine-wave--subtracted light curve shown by the black points in Fig.\ \ref{plt:0805} reveals that the long-term variations of the blazar needed to produce such an apparent phase shift are well within the variability characteristics of a normal blazar. So this type of phase shift in blazar SMBHB candidates is another likely phenomenological SMBHB blazar effect.

In Fig.~\ref{plt:0805phased}(a) we show the light curves from the four periods folded with a period of 1240.7 days, and in Fig.~\ref{plt:0805phased}(b) we show the same light curves, normalized to the same amplitude and the same offset from zero.

\section{Significance of the periodicity}\label{sec:significance}

To calculate the statistical significance of the observed periodicity in PKS~J0805$-$0111 discussed in \S~\ref{sec:analysis}, we followed the procedure outlined in O22. It  involves counting all simulations for which the $p$-value of the strongest peak is less than or equal to the $p$-value of the observed periodicity, the latter being the \textit{local} $p$-value, which is also estimated from the simulations. A summary of the procedure is given in Appendix \ref{app:sim}. In our analysis, we found that the period of $\sim 1245$ days has a global significance of $3.78$ $\sigma$ ($p$-value $=7.8 \times 10^{-5}$), when compared with  $1,000,000$ simulated light curves, with the best matching PSD and PDF to PKS~J0805$-$0111. The GLS peak power of these $1,000,000$ simulations and their periods are shown in Fig.~\ref{plt:glsp}. The result of the GLS analysis is shown in Table~\ref{tab:lsvalues}.

\section{The Effect of Sample Size}\label{sec:samplesize}

 A statistically complete, well-defined sample of 1158 blazars was tabulated by \citet{Richards2011} and was the initial sample in the OVRO monitoring program. This complete sample was selected from the Candidate Gamma-Ray Blazar Survey (CGRaBS), as introduced by \citep{Healey2008}.  The total CGRaBS sample consists of  1625 blazars  spread uniformly across the $|b| > 10^\circ$ sky, with flux densities $>65$ mJy  at 4.85 GHz, a radio spectral index $\alpha> -0.5$, and X-ray fluxes similar to those detected by the Energetic Gamma Ray Experiment Telescope (EGRET) detected sources.  
 The original OVRO 40 m Telescope monitoring sample included only the CGRaBS sources, numbering 1158,  above  declination  $\delta >-20^\circ$.
 Since then, many other objects have been added to the 40 m Telescope monitoring program and the whole sample now numbers $\sim 1830$.  These additional objects do not comprise a well-defined sample, so we  use only the initial sample of 1158 objects in our discussion below. This sample has the advantage that it was selected before the OVRO monitoring observations began, and so has not been influenced in any way by the results.

The global $p$-value that we derived for PKS J0805$-$0111 in \S~\ref{sec:analysis} is the probability, $p_0$, that a red noise process with the power-law PSD and PDF that best match the light curve of PKS~J0805$-$0111 produces a spurious GLS peak at any tested frequency that is locally at least as significant as the GLS peak deteced in PKS~J0805$-$0111. We derived a global $p$-value, $p_0=7.8 \times 10^{-5}$. 

It is clearly not the case that the 1158 objects in our well-defined sample all have the same variability characteristics as PKS~J0805$-$0111. However, we will assume, as a starting point, that PKS~J0805$-$0111 is representative of this sample as a whole, and that its PSD and PDF approximate the average of the sample. A detailed study will be carried out to characterize the variability properties of the whole sample, but this is not expected to change the results significantly, and is beyond the scope of this paper. We are confident that our present approach is adequate to give us the results we seek to the level of accuracy we seek.

 The probability of $k$ false positives in our sample of $n=1158$ is given by the binomial distribution. The probability ($p_{\rm PKS J0805-0111} (\geq 1)$)  that we get  at least one false-positive with $p-$value equal to, or less than, that found for  PKS J0805$-$0111 ($p = 7.8\times 10^{-5}$) in our sample, is 
\begin{equation}
p_{\rm PKS J0805-0111} (\geq 1)
= \sum_{k=1}^{\infty}\binom{n}{k} p^k (1-p)^{n-k} \approx 0.22\
\end{equation}

Thus there is  a substantial chance of finding  1 or more spurious periodicity detections with global p-value $p_0 \leq 7.8 \times 10^{-5}$  among a sample of 1158 sources with the same variability properties.  It is important to note, however, that, while   this means that the sinusoidal signal in PKS J0805$-$0111 \textit{could}  be due to red noise, it does mean that   it \textit{is}  due to red noise. O22  showed that the probability of the sinusoidally varying signal in PKS 2131$-$021 is due to red noise is $p=1.6 \times 10^{-6}$, i.e.  a $4.7\sigma$ result.  This means   that blazars exhibiting sinusoidal signals that are not due to red noise very likely exist in our sample.  We strongly recommend, therefore,  that the very few candidates in the OVRO 15 GHz monitoring program that pass the $3\sigma$ threshold we have set on the global GLS probability, should be counted as strong SMBHB candidates for follow-up with millisecond pulsar timing arrays unless or until proven otherwise.


\section{The Tip of the Iceberg}\label{sec:iceberg}

We have detected two blazars showing strong sinusoidal variations that dominate their radio light curves amongst 
blazars  in the OVRO  40\,m Telescope monitoring program.  It is important to recognize that, regardless of whether these candidates are SMBHBs or  not, these two examples are both exceptional in the relative strengths of their sinusoidal variations \textit{vs.} their random flux density variations.  It is almost surely the case, therefore, that there are many other blazars in this large sample with similar sinusoidal variations that do not dominate the light curve, and remain thus far undetected, so that the two cases we have reported thus far represent only ``the tip of the iceberg'', in other words  there must be many more such cases.   If a significant fraction of these blazars exhibiting sinusoidal variations are SMBHBs, then the high rate of occurrence will pose a challenge to theories of the formation of SMBHBs in galactic nuclei.

We are engaged in a comprehensive study of the 40 m Telescope light curves in which we add sinusoidal variations to the light  curves and then subject them to the GLS analysis described above in order to determine the probability of detecting the added sinusoid.  These tests are much complicated by the fact that the sinusoids can disappear and re-appear with different amplitudes.

Another aspect of the detection of two strong SMBHB candidates out of our monitoring sample of 1158 blazars, is that, if these candidates are confirmed as SMBHBs, then our two cases out of 1158 represent a lower limit to the number of SMBHBs in the sample. In the Kinetic Orbit model (\citealt{PKS2131II}, O22), the amplitude of the sinusoidal variability scales as $\cos i$ (O22), where $i$ is the inclination between the orbital angular momentum axis and the line of sight, so that, for geometric reasons, we may not detect some SMBHB candidates. The standard deviation of residuals from the best-fit sine-wave model (Fig.~\ref{plt:0805}) is 0.068 Jy. It is unlikely that we would be able to securely detect sinusoidal variations with an amplitude smaller than this. Assuming that the observed amplitude (0.171 Jy) is close to the maximum possible, we would not be able to detect the sinusoidal variations if $|i| > \textrm{arccos}(0.068/0.171) = 67^\circ$. If the orbits were distributed randomly, we will miss $(90-67)/90 \sim 25$\% of the SMBHs for simply geometric reasons. If we adopt a much more rigorous criterion for the detection of the periodicity (3$\sigma$ threshold), it is quite possible that we may miss more than 50\% of SMBHBs. If, in addition, we include the fact that the sinusoidal variations can appear and disappear on the timescale of one period, then it is clear that we will miss most of the SMBHBs in the monitoring sample,  For these reasons, we estimate that the fraction of SMBHB candidates amongst blazars is $\sim$ 1 in 100.

\begin{figure}[t]
   \centering
   \includegraphics[width=1.0\linewidth]{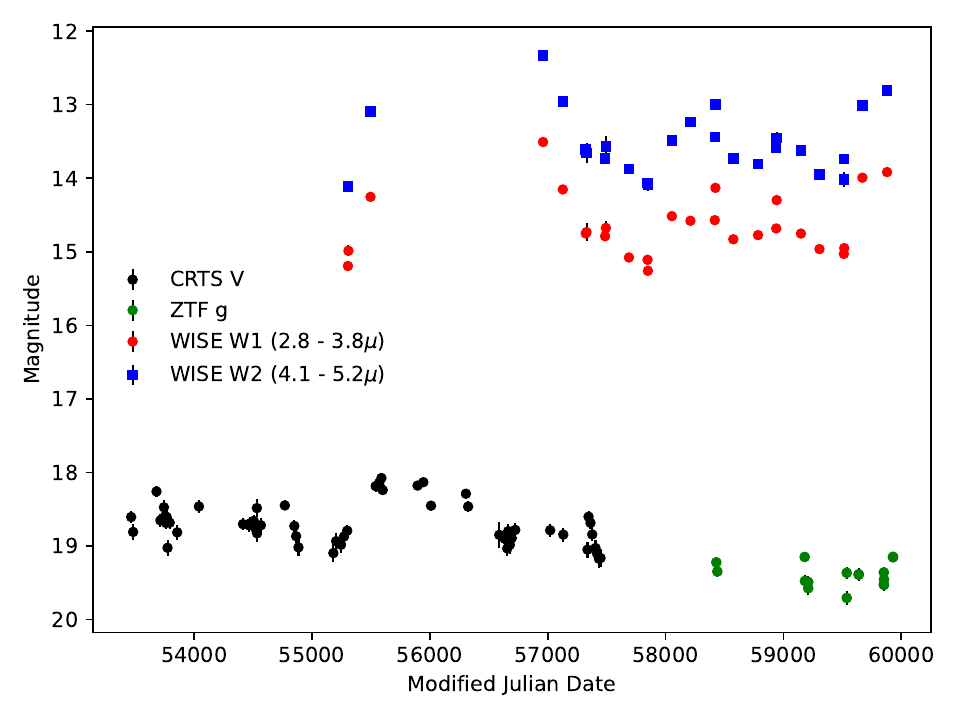}
      \caption{The archival optical and infrared light curve of PKS~J0805$-$0111 showing data from CRTS (black circles), ZTF $g$ (green circles), WISE W1 (red circles), and WISE W2 (blue squares). }
         \label{plt:0805_optir}
\end{figure}

 \section{Observations of PKS~J0805\texorpdfstring{$-$}{-}0111 at Infrared and Optical Energies}\label{sec:infrared}

The archival optical and infrared light curves of PKS~J0805$-$0111 are shown in Fig.~\ref{plt:0805_optir}, which includes 60 $V$-band data points from the Catalina Real-time Transient Survey \citep[CRTS;][]{Drake2009} taken with the 0.7m Catalina Schmidt Telescope located on Mt.~Bigelow, Arizona; and 33 $g$-band data points from the Zwicky Transient Facility \citep[ZTF;][]{Graham2019,Masci2019} at the Palomar 48-inch Oschin Schmidt telescope. The CRTS data were collected between 2005 May and 2015 December and the ZTF data between 2018 March and 2023 March, respectively. Data were also extracted for the blazar from the Wide Field Infrared Explorer (WISE) covering the 2.8--3.8$\mu$m (W1) and 4.1--5.2$\mu$m (W2) bands over the timeframe from 2010 April to 2022 October.

We have carried out a detailed cross-correlation analysis of the optical and infrared light curves with the OVRO 15\,GHz light curve of PKS~J0805$-$0111, identical to that on PKS~2131$-$021 described in O22, and we find no correlations above the $2\sigma$ threshold, so there is no evidence of significant correlations. However, it should be borne in mind that the optical and infrared data are sparse, so that this does not definitively rule out the possibility that there may be a correlation. 

 \section{Observations of PKS~J0805\texorpdfstring{$-$}{-}0111 at X-ray and \texorpdfstring{$\gamma -$}{gamma-}ray Energies}\label{sec:xand gamma}

Hard X-ray emission from blazars is important, because it is directly connected to the central engine, while soft X-rays from blazars are widely attributed to  synchrotron emission. The hard X-ray emission is attributed to the up-scattering of the soft photons of the synchrotron emission \citep{Mastichiadis2002PASA}, or relatively cool structures such as the accretion disk and the broad line region \citep{Dermer1993ApJ...416..458D,Sikora1994ApJS...90..923S}. While periodic oscillations are not uncommon in stellar mass BH X-ray Binary systems, such features are suprisingly rare in AGN X-ray observations \citep{gupta2018}. It is only very recently that such periodic signals have come to be widely reported and studied systematically \citep{Arcodia2021Natur,MiniuttiNatur2019}. In most cases, they originate from stellar mass compact objects interacting with the AGN accretion disk \citep{Franchini2023}. In the particular case of blazars, the X-ray light curves are usually chaotic and aperiodic. X-ray spectral-timing studies have revealed the existence of time-dependent flux and spectral state patterns. \citet{Bhatta2018A&A} explored the connections between blazar X-ray variability and other properties such as the spectral hardness and intrinsic X-ray flux in the \textit{NuSTAR} band. All 13 sources in their sample display high amplitude, rapid, aperiodic variability with a timescale of a few hours.

PKS J0805$-$011 has never been the main target of modern X-ray telescopes like \textit{Chandra} and \textit{XMM-Newton}. There are several \textit{Swift/XRT} observations but in many cases the off-axis angle is large such that the data are not very useful. We analyzed all \textit{Swift/XRT} data with off-axis angle smaller than 5'. The stacked spectrum reaches a total net exposure time of 8\,ks. The spectrum is characterized by a power-law of photon index $\Gamma\simeq2.2$, moderately absorbed by an intrinsic hydrogen column density of $8.5\times10^{21}\,\rm cm^{-2}$ in the rest frame ($z = 1.388$). The addition of an extra neutral Fe K$\alpha$ line improves the fit only marginally, though the observation is rather shallow. We conclude that the X-ray spectrum of PKS J0805$-$011 is well described by an absorbed featureless power-law. Future follow-up monitoring with large field-of-view X-ray instruments like \textit{Swift} and \textit{eRosita} will be promising, as simulations have predicted that close-separation accreting SMBHBs will have a periodically modulated hard X-ray component whose period is of order the binary orbital period \citep{Tang2018MNRAS,Krolik2019ApJ}.

\begin{figure}
   \centering
   \includegraphics[width=1.0\linewidth]{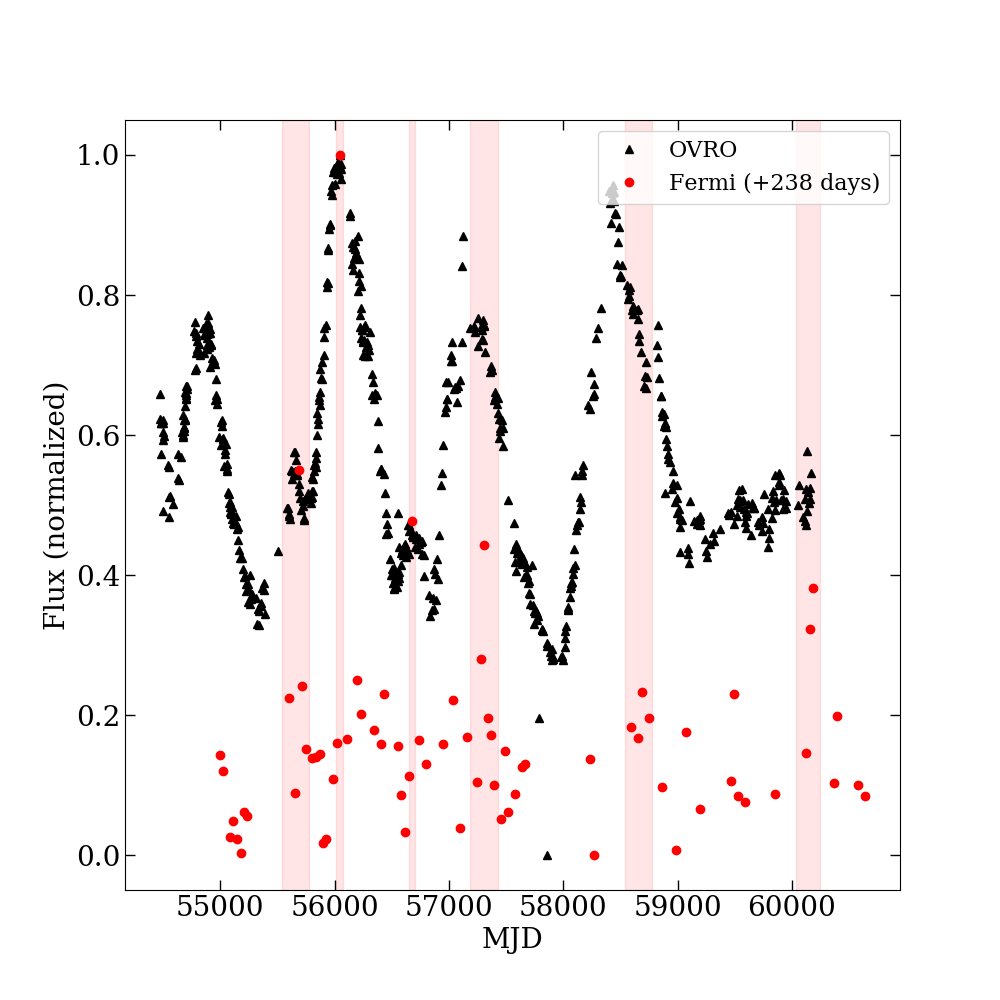}
      \caption{Normalized 15 GHz and GeV $\gamma$-ray light curve. The $\gamma$-ray light curve has been shifted forward in time by 238 days. The red shaded areas mark the periods of increased detection significance.}
         \label{plt:0805_fermi}
\end{figure}
PKS~J0805$-$0111  is in the top 26\% brightest $\gamma$-ray sources observed by the {\it Fermi} gamma-ray space telescope \citep{Abdollahi2020,Ajello2022}. It is included in the {\it Fermi} light curve repository\footnote{https://fermi.gsfc.nasa.gov/ssc/data/access/lat/LightCurveRepository/} \citep{Abdollahi2023} from which we extracted the 30-day binned (0.1-100~GeV) light curve shown in Fig. \ref{plt:0805_fermi}. We used the detection significance in each bin (test statistic -- TS) to identify periods of increased significance and hence higher flux. We then matched the highest flux observed in $\gamma$-rays to the highest flux density measured at 15 GHz. This requires shifting the $\gamma$-ray light curve forward in time by 238 days (MJD + 238 days).  Figure \ref{plt:0805_fermi} shows the normalized 15~GHz and $\gamma$-ray light curves shifted in time. It is evident that all the periods of increased $\gamma$-ray flux coincide with peaks in the radio emission.

\section{Gravitational-wave emission}\label{sec:grav}

If PKS~J0805$-$0111 is indeed a binary of supermassive black holes, it will produce gravitational waves with Earth-frame frequency $f_{\mathrm{GW}\Earth} = 2/P_\Earth \simeq 18.7$ nHz, within the sensitive range of pulsar timing arrays such as NANOGrav \citep{2023ApJ...951L...8A}, where $P_\Earth$ is the period that we observe in the light curve. For a quasi-circular orbit, the gravitational-wave strain is given by \citep{maggiore2008}
\begin{equation}
  h_{\mathrm{GW}} = \frac{2(\mathrm{G}{\mathcal{M}_\Earth})^{5/3}(\pi f_{\mathrm{GW}\Earth})^{2/3}}{\mathrm{c}^4 D_L},
  \label{eqn:strain}
\end{equation}
where $D_L \simeq 10$ Gpc is the luminosity distance and $\mathcal{M}_\Earth = (1+z) \times (M_1 M_2)^{3/5} / (M_1 + M_2)^{1/5}$ is the Earth-frame chirp mass, here given as a function of the rest-frame component masses $M_{1,2}$ and the redshift $z = 1.388$ \citep{Healey2008}.

By way of Eq.\ \eqref{eqn:strain}, upper limits on $h_{\mathrm{GW}}$ induce limits on the mass of the putative binary.
Specifically, taking the all-sky limit at $f_{\mathrm{GW}\Earth}$ derived from the 15-yr NANOGrav dataset \citep{2023ApJ...951L..50A} and calibrating it to the PKS~J0805$-$0111 sky location by way of the 6-nHz sky map (ibid.) yields a 95\% Bayesian upper limit $h_\mathrm{GW} \lesssim 10^{-14}$.
Under the assumption that the binary is approximately face on, this translates to rest-frame chirp mass $\mathcal{M} \lesssim 5 \times 10^9 \, M_\odot$, or to a limit $M \lesssim 11.4 \times 10^{9} M_\odot$ for the total rest-frame mass in a symmetric-mass binary. 
A joint constraint using electromagnetic data, as performed by \cite{2020ApJ...900..102A} for 3C 66B, could improve on these limits by a factor $\sim 2$.

As long as $\mathcal{M} \gtrsim 10^9 \, M_\odot$, the binary would be detectable with future pulsar timing arrays that include pulsars identified by the Square Kilometer Array \citep{2021ApJ...915...97X}.
However, as for PKS~2131$-$021, such high chirp masses imply short evolution times to binary coalescence ($\sim 2000 \, \mathrm{yr} \times (\mathcal{M} / 10^9 M_\odot)^{-5/3}$), casting doubt on the plausibility of observing such a massive late-stage system (O22).

\section{Conclusion}\label{sec:conclusions}

We emphasize that, in spite of the sample size, and the probability of $\sim 0.22$ of detecting a sinusoidal variation caused by the red-noise tail of the PSD in our sample, the fact that PKS~J0805$-$0111 exhibits all six of the likely characteristics of SMBHBs in blazars that we found in PKS~2131$-$021   shows that PKS~J0805$-$0111  should certainly be regarded as a strong SMBHB candidate.

The fact that the sinusoidal variations in PKS J0805$-$0111 have now disappeared, in a manner very similar to the disappearance of the sinusoidal variations in PKS 2131$-$021 in 1984, suggests a prediction based on the PKS 2131$-$021 behavior: namely -- that the sinusoidal variations will re-appear in PKS J0805$-$0111 after a gap of some years (in PKS 2131$-$021 the gap was 19 years), with the same period and in phase with the sinusoidal variations we are reporting here. In addition, continued monitoring all of the sources in the sample, including those that have hitherto shown no sign of periodicity, is very well-motivated. If and when more transitions are seen, multi-wavelength monitoring is also well-motivated.

We have now identified  two blazars, PKS~J0805$-$0111 and PKS~2131$-$021, that show strong sinusoidal variations having a global significance $p$-value $<1.3 \times 10^{-3}$, i.e. $> 3 \sigma$, which we interpret as real, and not due to random variations caused by the red-noise tail in the PSD.  The detection of a second strong SMBHB candidate showing sinusoidal variations demonstrates that PKS~2131$-$021 is not an isolated case.  Indeed, since we have picked out only blazars for which the light curves are dominated by the sinusoidal signal for most of the 40 m Telescope observing period, which must be the objects in our sample with orbits that are close to face on, there must be several candidates at significantly more oblique angles for every face on candidate we find, it is clear that $\sim$ 1 in 100 bright radio blazars must be SMBHB  candidates, which has interesting implications for searches for the SMBHB signal with pulsar timing arrays.


\section*{ACKNOWLEDGEMENTS}
We thank the California Institute of Technology and the Max Planck Institute for Radio Astronomy for supporting the  OVRO 40\,m program under extremely difficult circumstances over the last 5 years in the absence of agency funding. Without this private support this paper would not have been written.  We also thank all the volunteers who have enabled this work to be carried out.   Prior to 2016, the OVRO program was supported by NASA grants NNG06GG1G, NNX08AW31G, NNX11A043G, and NNX13AQ89G from 2006 to 2016 and NSF grants AST-0808050, and AST-1109911 from 2008 to 2014. The UMRAO program received support from NSF grants AST-8021250, AST-8301234, AST-8501093, AST-8815678, AST-9120224, AST-9421979, AST-9900723, AST-0307629, AST-0607523, and earlier NSF awards, and from NASA grants NNX09AU16G, NNX10AP16G, NNX11AO13G, and NNX13AP18G. Additional funding for the operation of UMRAO was provided by the University of Michigan. The NANOGrav project receives support from National Science Foundation (NSF) Physics Frontier Center award number 1430284. T. H. was supported by the Academy of Finland projects 317383, 320085, 322535, and 345899. S.K. and K.T. acknowledge support from the European Research Council (ERC) under the European Unions Horizon 2020 research and innovation programme under grant agreement No.~771282. I.L. and S.K. were funded by the European Union ERC-2022-STG - BOOTES - 101076343. Views and opinions expressed are however those of the author(s) only and do not necessarily reflect those of the European Union or the European Research Council Executive Agency. Neither the European Union nor the granting authority can be held responsible for them. W.M. acknowledges support by the ANID BASAL project FB210003 and FONDECYT 11190853. R.R. and B.M. and P.V.d.l.P. acknowledge support from ANID Basal AFB-170002, from Núcleo Milenio TITANs (NCN2023\_002), and CATA BASAL FB210003.
M.{}V.\ performed part of this work at the Jet Propulsion Laboratory, California Institute of Technology, under a contract with the National Aeronautics and Space Administration (80NM0018D0004).
V.P. acknowledges support from the Foundation of Research and Technology - Hellas Synergy Grants Program through project MagMASim, jointly implemented by the Institute of Astrophysics and the Institute of Applied and Computational Mathematics and by the Hellenic Foundation for Research and Innovation (H.F.R.I.) under the “First Call for H.F.R.I. Research Projects to support Faculty members and Researchers and the procurement of high-cost research equipment grant” (Project 1552 CIRCE). K.T. acknowledges support from the Foundation of Research and Technology - Hellas Synergy Grants Program through project POLAR, jointly implemented by the Institute of Astrophysics and the Institute of Computer Science. I. L. was supported by the NASA Postdoctoral Program at the Marshall Space Flight Center, administered by Oak Ridge Associated Universities under contract with NASA. The authors acknowledge the Kultrun Astronomy Hybrid Cluster (projects Conicyt Programa de Astronomia FondoQuimal QUIMAL170001, QUIMAL220002 and Conicyt PIA ACT172033) for providing HPC resources that have contributed to the research results reported in this paper.

\clearpage





\appendix
\twocolumngrid

\section{Simulations and Significance}
\label{app:sim}

\begin{figure}[t]
   \centering
   \includegraphics[width=1.0\linewidth]{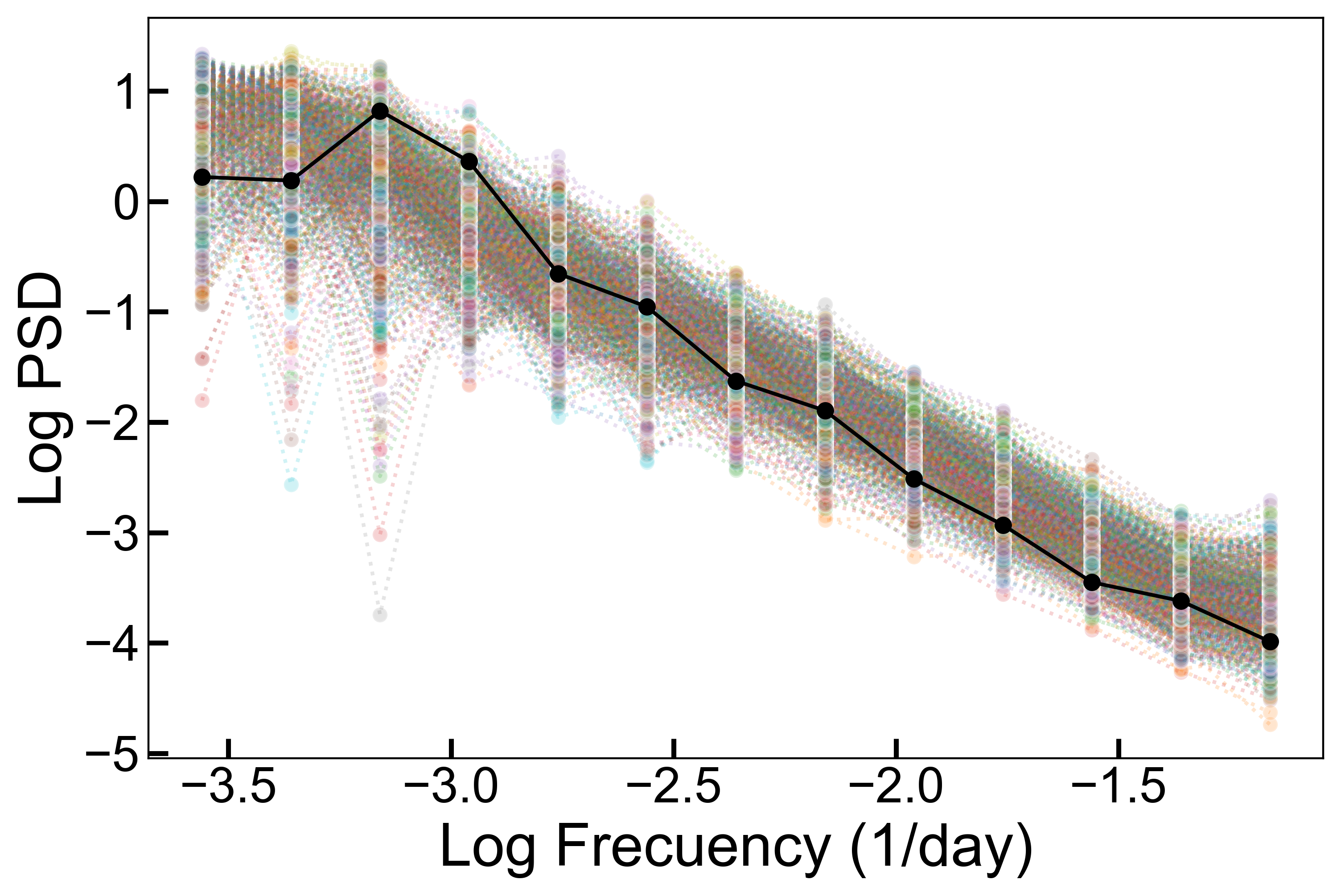}
      \caption{Power spectral density (PSD) over frequency. Black dots and line: the observed PSD of PKS~J0805$-$0111. Coloured dots and lines: PSDs of 1,000 randomly selected simulations.}
         \label{plt:0805dist_psd}
\end{figure}

For this study, we simulated light curves which best match the PSD, the PDF, sampling, and observational noise of the observed light curve under the assumption of a power law red noise process. Specifically, we assumed that the PSD $ = \textrm{constant} \times \nu^{-\beta} + P_\mathrm{white}$, where $\nu$ is the frequency. We followed a procedure based on  \citet{uttley2002} and \citet{2014MNRAS.445..437M} to obtain the slope which best matches the observed PSD, in the case of PKS~J0805$-$0111 $\beta = 1.85 \, [1.75$--$2.083]$, see Fig.~\ref{plt:0805dist_psd}.

A noise floor level $P_\mathrm{white}$ was added to account for the flux-density uncertainties. To calculate the contribution of $P_\mathrm{white}$ we assumed white noise $\mathcal{N}(0, \sigma_e^2)$ where $\sigma_e^2$ is the mean variance of the uncertainties. Then, using Parseval's Theorem, we estimated the contribution of $P_\mathrm{white}$:
\begin{align*}
    \sigma_e^2 &= \int_{\nu_0}^{\nu_1} P(\nu) d\nu = P_\mathrm{white}\int_{\nu_0}^{\nu_1} d\nu \\ 
    &= P_\mathrm{white} (\nu_1 - \nu_0),
\end{align*}
where $\nu_0$ is the lowest sampled frequency and $\nu_1$ is the Nyquist frequency. 

Note that the second and third point of the observed PSD in Fig.~\ref{plt:0805dist_psd} deviate from a general linear trend. This increased power over the typical power law red noise spectrum at these frequencies is the contribution of the periodic signal. The simulations implement the pure red noise hypothesis and therefore do not systematically reproduce this peak.

\begin{figure}[t]
   \centering
   \includegraphics[width=\linewidth]{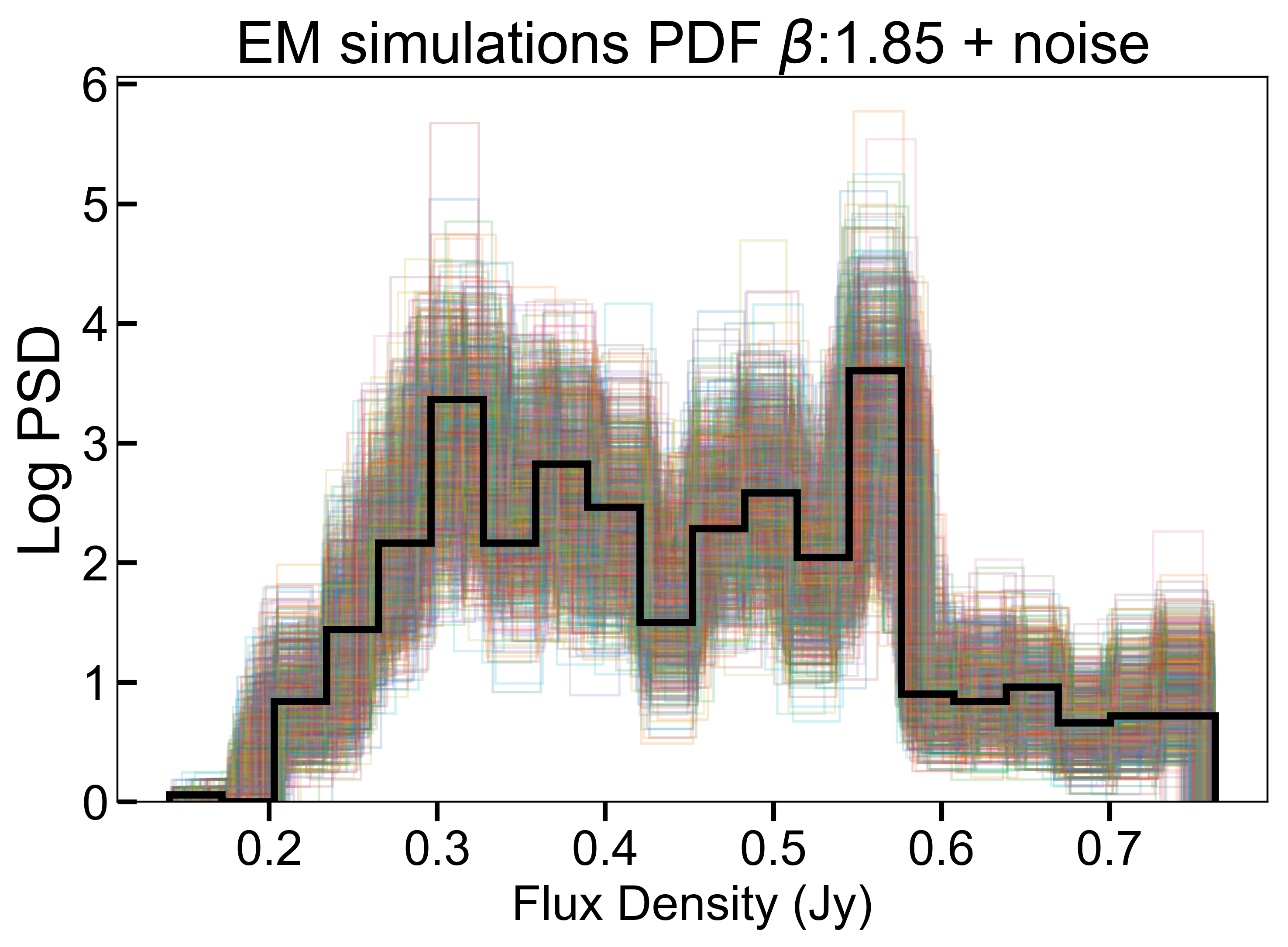}
      \caption{Probability density function (PDF). Black histogram: the observed PDF of PKS~J0805$-$0111. Coloured dots and lines: PDFs of 1,000 randomly selected simulations.}
         \label{plt:0805dist_PDF}
\end{figure}

Considering the non-Gaussian PDF, see Fig.~\ref{plt:0805dist_PDF}, we adopted the algorithm introduced by \citet{2013MNRAS.433..907E} for our simulations. This algorithm allows us to create simulations with the above PSD and a PDF matching that of the observed light curve. For characterizing the PDF we used the empirical cumulative distribution function. Simulated lightcurves 10~times longer than the observed lightcurve and with 10~times higher cadence were resampled to match the original sampling pattern.
The PDF of the simulations matches the observed PDF, see Fig.~\ref{plt:0805dist_PDF}.

After generating the artificial light curves, we carried out a rigorous significance analysis of periodicity detections through the following steps:

\begin{enumerate}
    \item We calculated the power spectra of observed data and simulated light curves using the GLSP. For this study, the number of simulations is $1,000,000$.
    \item We compared the strongest observed peak, at its specific period, with the ensemble of simulated GLSPs. Specifically, we counted the fraction of simulations where the GLSP power is equal or greater than the observed peak at the same period. This gives us the \emph{local $p$-value} $p_\mathrm{peak}$, i.e. evaluated at the peak period. To account for the \emph{look elsewhere effect} of analysing a large number of frequencies we carried out two additional steps.
    \item In the same fashion as for the observed data, we identified the strongest peak in the GLSP of each simulation and calculated its local $p$-value.
    \item Finally, we counted the fraction of simulations, for which the local $p$-value $p_\mathrm{sim}$ is equal to or lower than the observed local $p$-value. This gave us the \emph{global $p$-value}.   
\end{enumerate}

The local $p$-value can only be used if the period is known a priori  from independent tests. If that is not the case, as in this case, we have to use the global $p$-value.
We use a conservative threshold of $p$-value $<1.3 \times 10^{-3}$, equivalent to a significance level $>3\sigma$, to reject the hypothesis that the periodic signal arises spuriously from a red-noise process.

\section{Subtracted Sine Wave}

The significance analysis we described above is based on the hypothesis that the radio light curve of PKS~J0805$-$0111 follows a red-noise process with a power-law PSD with power-law index $\beta = 1.85 \, [1.75$--$2.083]$ estimated from the spectrum shown in Fig.~\ref{plt:0805dist_psd}, which gives us the best matching purely stochastic model of the light curve. We rejected this model with a global $p$-value of $7.8 \times 10^{-5}$ at the $3.78\sigma$ significance level.
However, Fig.~\ref{plt:0805dist_psd} shows that the periodic signal is visible in the PSD as increased power at the third and fourth lowest tested frequencies. This may bias our power-law index estimation towards steeper spectra. 

In order to verify that a possible bias in the PSD index estimation does not affect our conclusion, we carried out a second analysis using the residual light curve after subtracting the sinusoidal component.
The black curve in Fig.~\ref{plt:0805} shows the residual light curve. From these data we estimated a PSD power-law index $\beta= 1.6 [1.5$--$1.7]$, corresponding to flatter PSD than estimated from the original data. We repeated the entire periodicity analysis with simulations that have $\beta= 1.6$ and found a global $p$-value of $2.1 \times 10^{-4}$, which, which differs by only a factor 2.7 from our first result, and therefore has no effect on our findings. We conclude that our results do not depend on the specific PSD slope value and are not affected by any potential bias that a spurious or intrinsic period may introduce into the PSD estimate.

\begin{figure}[t]
   \centering
   \includegraphics[width=1.0\linewidth]{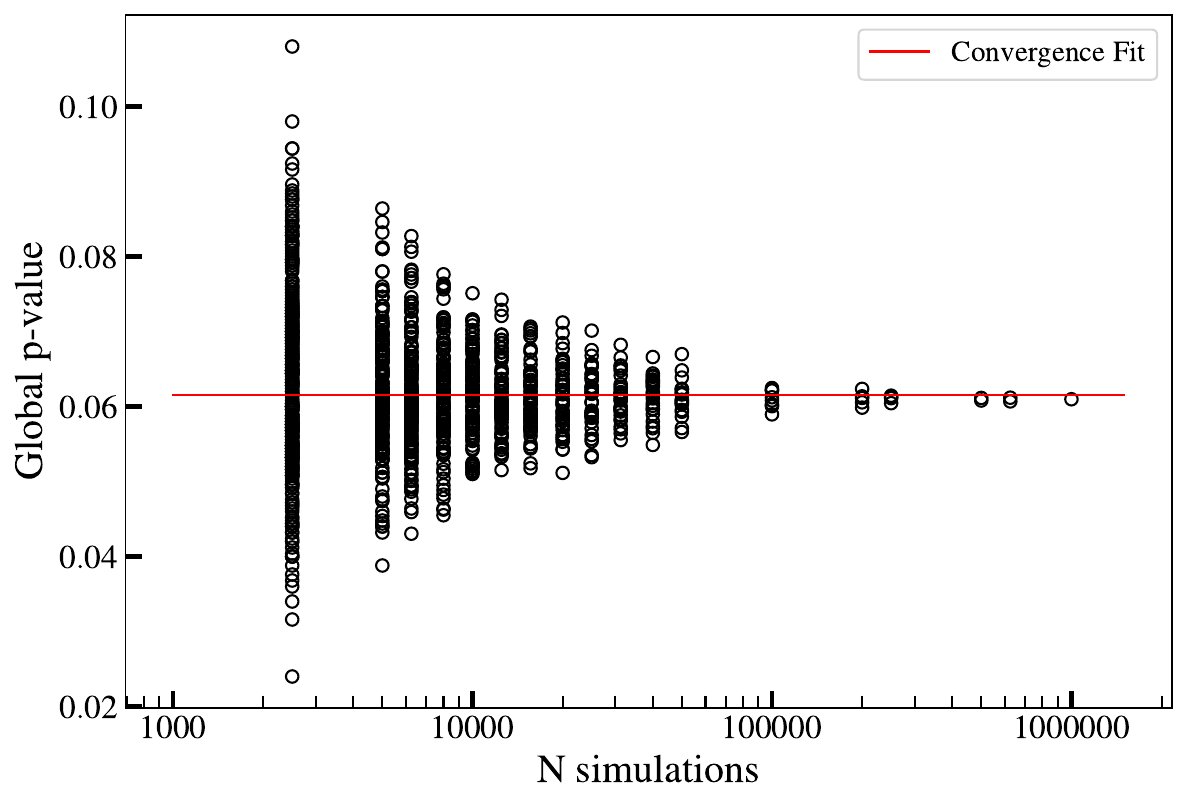}
      \caption{Convergence of global $p$-value with increasing number of simulations when the local $p$-value is $>0$. Based on the OVRO 40\,m and the strongest peak in residual GLSP data of PKS~J0805$-$0111 with sine-wave subtracted.}
    \label{plt:pvaluelinear}
\end{figure}

\begin{figure}[t]
   \centering
\includegraphics[width=1.0\linewidth]{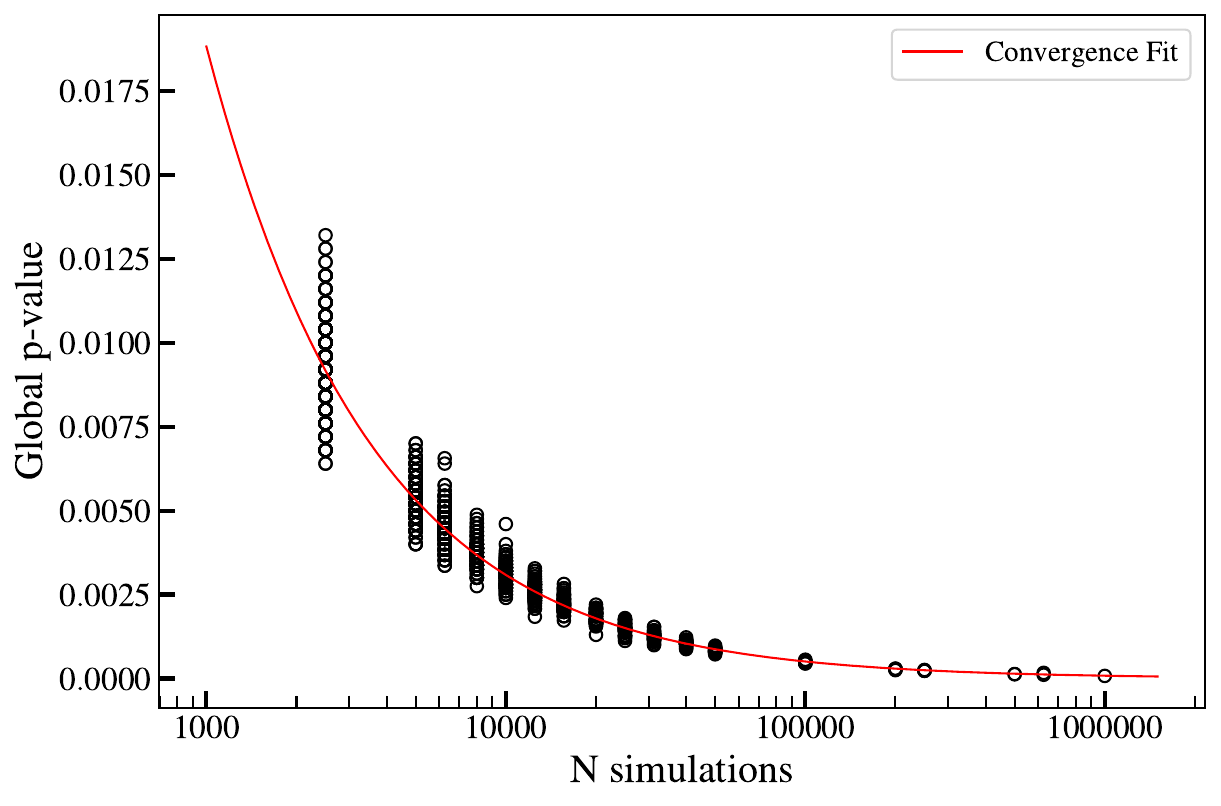}
      \caption{Convergence of global $p$-value with increasing number of simulations when the local $p$-value is $0$. Based on the OVRO 40\,m data of PKS~J0805$-$0111 with sine-wave subtracted.}
    \label{plt:pvaluedecrease}
\end{figure}

\section{Convergence of the global \texorpdfstring{$p$}{p}-value}

It is important to estimate the accuracy of global $p$-value, which depends on the total number of simulations. We methodically generated different subsets from our 1,000,000 simulations. From each subset we calculated the global $p$-value. We need to distinguish two situations that depend on the local $p$-value:

\begin{itemize}
    \item \textit{Case~I: local $p$-value $>$ 0.}    Fig.~\ref{plt:pvaluelinear} shows how the global $p$-value and its accuracy depend on the number of simulations, based on the OVRO 40\,m data of PKS~J0805$-$0111 with sine-wave subtracted, where the residuals generate a peak at a period of 2445 days with a local $p$-value of $\sim 1.28 \times 10^{-2}$ and a global $p$-value of $\sim 1.42 \times 10^{-1}$.
    When the local $p$-value is well defined, increasing the number of simulations reduces the scatter of the estimated global $p$-value around the true value. The larger scatter at fewer simulations arises from the stochastic nature of the simulations. The reduction in the scatter indicates that we reached adequate accuracy with 1,000,000 simulations in this particular case; the behaviour of the convergence can be seen in Fig.~\ref{plt:pvaluelinear}.
    \item \textit{Case~II: local $p$-value equal 0.} 
    Fig.~\ref{plt:pvaluelinear} shows how the global $p$-value and its accuracy depend on the number of simulations, based on the OVRO 40\,m data of PKS~J0805$-$0111 as discussed in this paper.
    A local $p$-value equal to 0 means that none of our 1,000,000 simulations have a a GLS power at least as strong as the power of the GLSP of the observed light curve at the peak period, $P_\mathrm{peak}$. Therefore, the local $p$-value is not well estimated, but rather has an upper limit of $<1/1,000,000$. Accordingly, a smaller subset of simulations results in a higher upper limit on the local $p$-value, which then results in a higher upper limit of the global $p$-value. Consequently, the estimate of the global $p$-value decreases with increasing number of simulations and converges towards the true value. In this case the global $p$-value is biased towards being over-estimated and rather poses an upper limit. Similar to case~I, increasing the number of simulations also reduces the scatter in the estimate of the $p$-value, or more accurately its upper limit.
    In the case of PKS~J0805$-$0111, Fig.~\ref{plt:pvaluedecrease} indicates that we have reached adequate accuracy in the estimate of the $p$-value and we have passed our $3\sigma$-confidence limit with 1,000,000 simulations. Whereas further increasing the number of simulations would give us a more accurate estimate of the $p$-value, it would not change our conclusions.
\end{itemize}


\bibliographystyle{aa.bst}
\bibliography{aanda}

\end{document}